# BS-GAT：Behavior Similarity Based Graph Attention Network for Network Intrusion Detection

Yalu Wang[1], Zhijie Han[2*], Jie Li[3], Xin He[4]

[1]School of Computer and Information Engineering, Henan University, Kaifeng 475004, China.

[2]International Joint Laboratory of Intelligent Network Theory and Key technology of Henan, Henan University, Kaifeng 475004, China.

[3]School of Intelligent Engineering, Zhengzhou University of Aeronautics, Zhengzhou 450046, China.

[4]School of Software, Henan University, Kaifeng 475004, China.

**\* Correspondence:**
Zhijie Han
hanzhijie@126.com

## Abstract

With the development of the Internet of Things (IoT), network intrusion detection is becoming more complex and extensive. It is essential to investigate an intelligent, automated, and robust network intrusion detection method. Graph neural networks based network intrusion detection methods have been proposed. However, it still needs further studies because the graph construction method of the existing methods does not fully adapt to the characteristics of the practical network intrusion datasets. To address the above issue, this paper proposes a graph neural network algorithm based on behavior similarity (BS-GAT) using graph attention network. First, a novel graph construction method is developed using the behavior similarity by analyzing the characteristics of the practical datasets. The data flows are treated as nodes in the graph, and the behavior rules of nodes are used as edges in the graph, constructing a graph with a relatively uniform number of neighbors for each node. Then, the edge behavior relationship weights are incorporated into the graph attention network to utilize the relationship between data flows and the structure information of the graph, which is used to improve the performance of the network intrusion detection. Finally, experiments are conducted based on the latest datasets to evaluate the performance of the proposed behavior similarity based graph attention network for the network intrusion detection. The results show that the proposed method is effective and has superior performance comparing to existing solutions.

**Keywords: Internet of Things[1], Network intrusion[2], Graph Neural Network[3], Behavior similarity[4], Graph construction[5].**

## 1    Introduction

With the rapid development of the Internet of Things (IoT) in recent years, IoT devices have undergone significant growth in scale. For example, they have a very high degree of complexity and different devices having different functionalities, such as network cameras, smart TVs, smart curtains, wireless temperature sensors, wireless printers, and smart kitchen appliances. These devices are interconnected to form edge devices in the network[1]. Therefore, attacks on IoT networks, in terms of both quantity and frequency, are increasing exponentially. To promote the development of IoT and maintain the security of the intelligent networks, it is important to research on network intrusion detection methods in IoT environments.

There are mainly two types of network intrusion detection methods: feature-based detection and machine learning-based detection[2]. Feature-based intrusion detection systems require a pre-set group of attack features to compare and detect the captured data flows. The accuracy of this detection method depends on the quality of the features, and it cannot react well to new types of attacks, resulting in poor detection performance. With the development and application of machine learning in recent years, machine learning-based intrusion detection methods have emerged. Most machine learning-based methods directly use data flow information for identification[3][4][5][6][7]. Although these methods can effectively solve the feature dependence problem and have good detection performance on new attacks, their identification accuracy is too low to apply them to practical networks. Furthermore, in recent years, researchers have focused on the relatively new subfield of machine learning, graph neural networks (GNNs), and proposed some new intrusion detection methods[2][8][9]. GNNs are a new type of neural network developed for the correlation properties of graphs. In recent years, they have been extensively applied in areas such as graph processing[10], networks[11], intelligent transportation[12], recommendation systems[13], and distributed computing[14]. Because networks are natural graphs, it is appropriate to introduce graph neural networks into network intrusion methods in order to effectively improve detection accuracy.

The GNNs based intrusion detection method consists of two steps: graph construction and neural network identification. Graph construction refers to the process of using existing network information and mapping relationships to generate a logically consistent graph. The neural network identification refers to the method of using a neural network on a generated graph to identify nodes or edges. Existing GNNs based intrusion detection methods have improved the detection performance comparing to traditional machine learning based intrusion detection methods. However, there are practical issues in the graph construction and neural network identification steps for existing GNNs based intrusion detection methods. For example, the logic of graph construction is very simple and does not consider the characteristics of network data flow[2][9]. The direct use of existing graph neural network methods for identification does not consider the relationships between nodes[8]. These problems result in low identification accuracy for GNNs based intrusion detection methods, especially in multi-classification problems. Therefore, it is urgent to further investigate the GNNs based intrusion detection problem by considering the above issue.

This paper proposes a graph neural network algorithm based on behavior similarity (BS-GAT) using graph attention network for network intrusion detection. Both the graph construction and the neural network identification are improved based on the behavior similarity. In the step of the graph construction, the data flows are represented as nodes on the graph, and the behavior rules are represented as edges to construct the graph with a relatively uniform number of neighbors for each node. In the step of the eural network identification, the behavior similarity is introduced to adapt the characteristics of the practical datasets. Then, the graph attention network (GAT) is introduced, to propose a behavior similarity-based graph attention network (BS-GAT) for network intrusion detection. To utilize the relationship between data flows and the structure information of the graph, the edge behavior relationship weights are incorporated into the GAT network to calculate the features of the hidden layer nodes.

In summary, this paper has three main contributions:

- We investigate the network intrusion detection problem by proposing a novel behavior similarity based graph attention network. A novel graph construction method is developed using the behavior similarity, The data flows are treated as nodes in the graph, and the behavior rules of nodes are used as edges in the graph, constructing a graph with a relatively uniform number of neighbors for each node.

- We incorporate the edge behavior relationship weights into the GAT network to calculate the features of the hidden layer nodes. This enables a better utilization of the relationship between data flows and the structure information of the graph for network intrusion detection methods.

- The proposed intrusion detection method is evaluated using experiments based on the latest network flow datasets. By comparing with graph neural network-based intrusion detection methods and traditional intrusion detection methods, the experimental results show the superier performance of the proposed method.

The rest of the paper is organized as follows. Section 2 introduces the research progress of scholars in the field of network intrusion detection methods. Section 3 first introduces some information about GNN and GAT, and then elaborates on the proposed method in detail. Section 4 describes the datasets, experimental methods, and results. Section 5 provides a summary of this paper.

## 2   Related Work

In recent years, many researchers have developed machine learning-based methods for network intrusion detection. However, most of these methods directly use data flow information for identification.

Haitao He et al.[3] extracted different levels of features from network connections. And further, they proposed multimodal-sequential intrusion detection method with special structure of hierarchical progressive network, which is supported by multimodal deep auto encoder (MDAE) and LSTM technologies. Then, experiments were conducted on the NSL-KDD, UNSW-NB15, and CICIDS 2017 benchmark datasets, achieving an accuracy of 94% and 88% in binary and multi-class classification problems, respectively.

In[6], the authors focused on securing Internet of medical things' networks (IoMT) by using a two-level intrusion detection model. The first level used a decision tree (DT), naive Bayes (NB), and random forest (RF) as first-level individual learners. In the next level, an XGBoost classifier was used to identify normal and attack instances. The ensemble model achieved a binary classification accuracy of 96.35% and an F1-Score of 0.95 on the ToN-IoT dataset.

The two methods mentioned above use traditional machine learning techniques for network intrusion detection, such as decision trees, MDAE, LSTM, random forests, and XGBoost. These methods directly train on the dataset without considering the graph topology information, resulting in each training data being relatively singular and unable to fully explore the information in each data. Therefore, these methods have limited capability in detecting complex network attacks, such as Botnet attacks[15], distributed port scans[16] or DNS Amplification attacks[17],which require a more global network view and traffic.

Qingsai Xiao et al.[18] converted network traffic into first-order and second-order graphs by combining statistical features with latent features. The first-order graph learns the latent features from the perspective of a single node, and the second-order graph learns the latent features from a global perspective. Then, the traditional machine learning classifier method was used for anomaly detection, which has good results on two real data sets. However, this model is a direct push, so it may not be able to classify unknown network flows. And it relies on the originally formed graph, which is not friendly to extensibility.

Due to the good performance of GNN on graph-structured data and the fact that a network is a natural graph, in recent years, some scholars have applied GNN to intrusion detection systems.

Cheng Q et al.[8] proposed an Alert-GCN framework that uses graph convolutional networks (GCN) to associate alerts coming from the same attack for the prediction of the alert system. Then adopted the DARPA99 data set for experiments, achieving good results in the experiment. However, this method used a custom similarity method to define the edges in the graph when constructing an alarm-related graph, which will cause the neighbors of any node in the graph to be highly similar to themselves. This paper recognizes that the graph constructed by this method will cause overfitting in the graph neural network.

Wai Weng Lo et al.[2] proposed the E-GraphSAGE architecture based on GraphSAGE, which allows capturing the edge features and topology information of the graph for network intrusion detection in the Internet of Things. Experiments on four latest NIDS benchmark datasets show that the method has achieved very good effects on binary classification problems, but not on multi-classification problems. When constructing the graph, this method directly uses the ip address and port of the network flow as nodes and the network flow as edges. When this method uses GraphSAGE for neighborhood aggregation, it may cause some neighborhoods to be aggregated multiple times and the whole neighborhood to be aggregated, which will affect the identification accuracy.

Chang, Liyan et al.[9] improved the E-GraphSAGE architecture and proposed an E-ResGAT architecture based on GAT. The key idea is to integrate residual learning into GNN by using available graph information. Residual connections were added as a strategy to deal with the high class imbalance, aiming at retaining the original information and improving the minority classes' performance. An extensive experimental evaluation on four recent intrusion detection datasets shows the excellent performance of the approaches, especially when predicting minority classes. However, the graph construction method of this scheme is improved on the basis of E-GraphSAGE, converting edges to nodes and nodes to edges through bipartite graph. This method still cannot eliminate the problem of E-GraphSAGE graph construction.

Based on the above, it is crucial to construct a logical graph when using GNN for intrusion detection. Therefore, inspired by the self-similarity of network traffic[19], this paper proposes a new graph construction method based on the principle of behavioral similarity[20], and then assigns behavioral weights to the edges of the generated graph, which are applied to the aggregation operation of GAT. Finally, the effectiveness of this method is demonstrated in experiments.

## 3 Behavior Similarity based GAT for Network Intrusion Detection

### 3.1 GNN and GAT

GNN is one of the rapidly developing subfields in the field of deep learning in recent years. Its strength lies in its strong ability to learn from graphs, enabling it to capture information hidden in graphs. Additionally, it has the ability to generalize to dynamic changes in network topology, making it very friendly to non-spectral graph-structured data. With the development of GNN, many derivatives have also emerged. Among them, GAT[21], GCN[22], GraphSAGE[23], etc. have received considerable attention from scholars. Considering that GAT is a graph neural network that introduces attention mechanism, it can automatically calculate the attention coefficient between two nodes through learning, and aggregate the information of node neighbors through attention coefficients. Therefore, this type of graph neural network is more capable of mining graphical information. This paper applies GAT to network intrusion detection methods.

GAT has two core components: computing the attention coefficients $\alpha_{ij}$ and the hidden layer features $\vec{v}_i'$ of nodes.

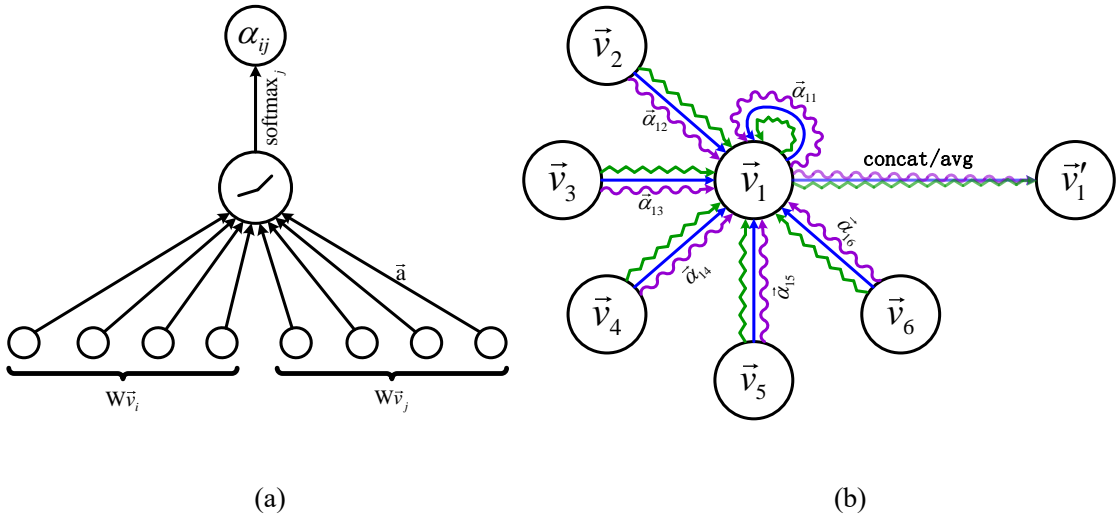

(a)                  (b)

Figure 1: (a): The calculation process of attention coefficient. (b): The calculation process of hidden layer features.

The calculation process for the attention coefficients $\alpha_{ij}$ between neighboring nodes is shown in Figure 1(a). The original GAT paper mentions two types of neighborhoods: node neighbors (nodes directly connected to the current node) and full neighborhood (all nodes). In this paper, we use node neighbors as the neighborhood to prevent the aggregation of a large amount of neighborhood node information that may cause the features to become blurry. Using node neighbors can also reduce time and space complexity.

$$\alpha_{ij} = \frac{\exp\left(\text{LeakyReLU}(\vec{a}^T[W\vec{v}_i||W\vec{v}_k])\right)}{\sum_{k\in\mathcal{N}_i}\exp\left(\text{LeakyReLU}(\vec{a}^T[W\vec{v}_i||W\vec{v}_k])\right)} \quad (3)$$

$a$ is a single-layer feedforward neural network, which is parameterized by a weight vector $\vec{a} \in \mathbb{R}^{2F'}$, $W \in \mathbb{R}^{2F' \times F}$ is a shared linear transformation matrix, LeakyReLU is a nonlinear function, and $\mathcal{N}_i$ is the neighbor node of node $i$. The equation finally uses a $softmax$ function for normalization.

The calculation process of hidden layer features $\vec{v}_i'$ is shown in Figure 1(b). The calculation equation is:

$$\vec{v}_i' = \sigma(\frac{1}{K}\sum_{k=1}^{K}\sum_{j \in \mathcal{N}_i} \alpha_{ij}^k W^k \vec{v}_j) \tag{4}$$

$\sigma$ is a nonlinear function. Since GAT uses a multi-head attention mechanism, K represents the number of attention heads.

### 3.2 BS based Graph Construction

The first step in using a GNN is to construct a logically consistent graph. When constructing a graph, we hope that the graph is uniform and that the number of neighbors of nodes is balanced, as shown in Figure 2(a). If a node has too many neighbors, pooling information using a graph neural network can cause the node's features to become blurry, leading to misidentification. If a node has too few neighbors, the node may become an isolated point or edge. For example, the dashed ellipse in Figure 2(b) shows an isolated edge with no neighbors, making it impossible for the graph neural network to extract any meaningful information. Some graph neural network-based network intrusion detection methods, such as E-GraphSAGE and E-ResGAT, construct graphs that are uneven.

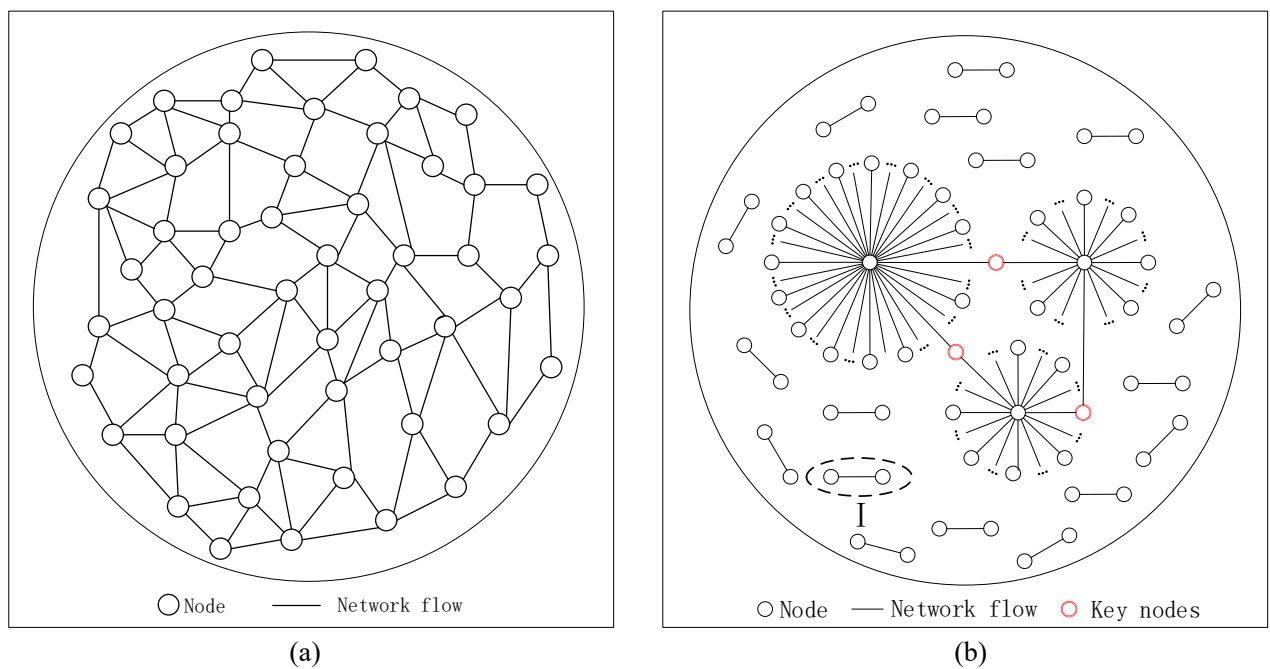

Figure 2 (a): The expected constructed graph. (b): Graphs constructed by other algorithms.

IoT devices are connected through networks, and the network flow data between IoT devices is equivalent to device behavioral data. IoT devices can be seen as a person, and network flow data is equivalent to human behavior. Network flow identification is equivalent to social network persona profiling.

Human behavior similarity has three characteristics:

1) There is self-similarity between different behaviors of a person.

2) Behaviors among individuals of the same group are highly similar, but with some differences.

3) Behaviors among different groups of people exhibit significant differences.

The three points mentioned above can be mapped to the behavior of IoT nodes. If the source IP addresses of data flows are the same, they are classified as the behavior of one node. If the subnet masks are the same, they are classified as the behavior of the same type of node. Therefore, the process of graph construction can be divided into the following steps:

**Step1:** All data flows are set as nodes of the graph $\mathcal{V} = \{\vec{v}_1, \vec{v}_2, \ldots, \vec{v}_n\}$, as shown in Figure 3(a), where $\vec{v}_i$ is the feature vector of node $i$ and $n$ is the number of nodes.

**Step2:** According to point 1) of behavior similarity, connect all nodes with the same source IP address in pairs to generate an edge $e_{ij}^s$, $(i, j \in n \mid e_{ij}^s \in \mathcal{E}^s)$, and generate self-similar areas, such as Figure 3(b).

**Step3:** According to point 2) of behavior similarity, nodes with the same subnet mask are classified into one category, and edges are constructed between self-similar areas in Step 2. Then, the mask areas are generated, as shown in Figure 3(c). The edge construction rule is when two nodes have accessed the same destination IP address on the same port, the edge $(e_{ij}^m, (i, j \in n \mid e_{ij}^m \in \mathcal{E}^m))$ is constructed between the two nodes.

**Step4:** According to point 3) of behavior similarity, build the edge between the mask areas in step 3, as shown in Figure 3(d). First, each node searches for the mask area with the longest common prefix with itself, and then searches for nodes with the same source port, destination IP address, and destination port within this area, and then the edge $(e_{ij}^o, (i, j \in n \mid e_{ij}^o \in \mathcal{E}^o))$ is constructed between the two nodes.

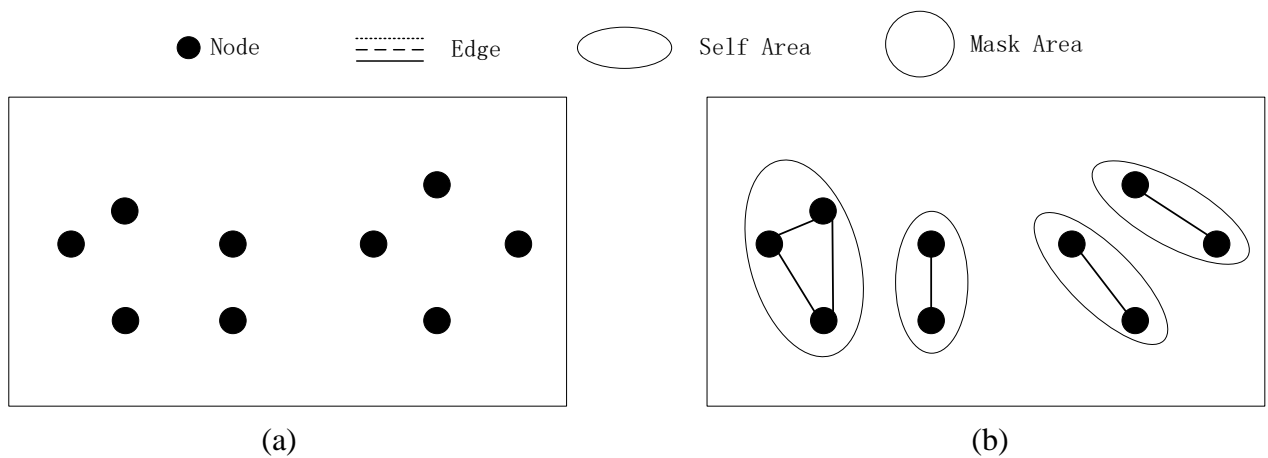

(a)         (b)

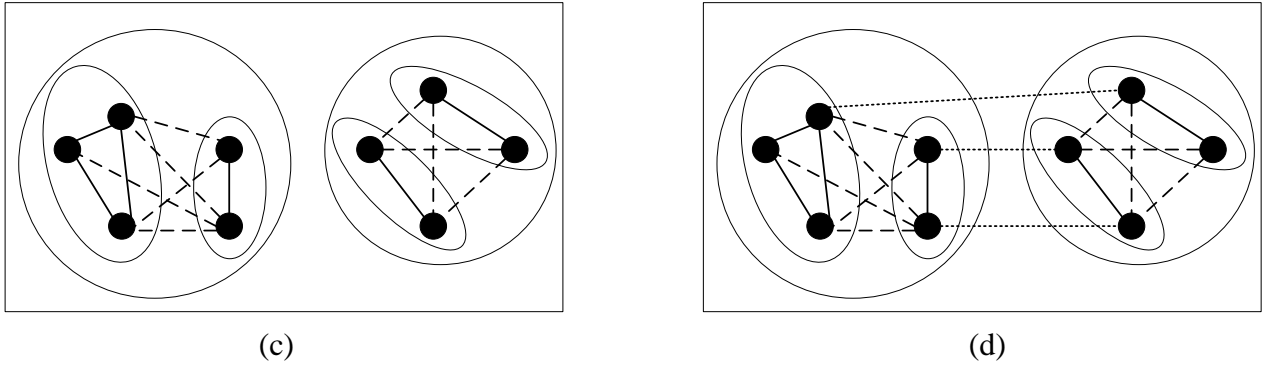

(c)　　　　　　　　　　　　　　　　(d)

Figure 3: Graph construction method

Through the four steps above, the node set $\mathcal{V}$ and the edge set $\mathcal{E}$ are obtained. The two sets form a graph $\mathcal{G}(\mathcal{V}, \mathcal{E})$, where E is represented as follows :

$$\mathcal{E} = \mathcal{E}^s \cup \mathcal{E}^m \cup \mathcal{E}^o \tag{1}$$

Then it is necessary to construct a behavior weight of all edges, which aims to balance the attention mechanism of the GAT network. The reason will be explained in the next section. The behavior weight of the edge is expressed as:

$$eb_{ij} = \begin{cases} 1 & , e_{ij} \in \mathcal{E}^s \\ \lambda & , e_{ij} \in \mathcal{E}^m \\ \mu & , e_{ij} \in \mathcal{E}^o \\ 0 & , other \end{cases} \quad i, j \in n \tag{2}$$

$\lambda$ represents the weight value of the edge between the self-similar areas, and $\mu$ represents the weight value of the edge between the mask areas. The relationship between λ and μ that needs to be explained is $1 < \lambda < \mu < 0$, making it as a hyperparameter for neural network training. Its value cannot be set too small to avoid having a significant impact on training. In this paper, $\lambda$ is set to 0.85 and $\mu$ to 0.7.

Through the above methods, we write the pseudocode as follows:

---
Algorithm 1: Graph Construction
---
Input: node features $\vec{v}_i$;

Output: graph $\mathcal{G}(V, E)$; behavior weight of edge $eb_{ij}$;

---
1　$V \leftarrow \vec{v}$
2　**for** $i = 1, \ldots, n$ do　　// $n$ is the number of nodes
3　　**for** $j = i, \ldots, n$ do
4　　　**if** $\vec{v}_i$ and $\vec{v}_j$ have the same source IP address
5　　　　$E = E \cup e_{ij}$
6　　　　$eb_{ij} = 1$
7　　　　**continue**
8　　**end**
9　　　**if** $\vec{v}_i$ and $\vec{v}_j$ are in the same subnet and have the same destination IP address and

|     | destination port |
| --- | --- |
| 10  | $E = E \cup e_{ij}$ |
| 11  | $eb_{ij} = \lambda$ |
| 12  | **continue** |
| 13  | **end** |
| 14  | **if** $\vec{v}_i$ and $\vec{v}_j$ have the longest prefix in the subnet mask and have the same source port, destination port, destination IP address |
| 15  | $E = E \cup e_{ij}$ |
| 16  | $eb_{ij} = \mu$ |
| 17  | **continue** |
| 18  | **end** |
| 19  | **end** |
| 20 **end** | |
| 21 $\mathcal{G} \leftarrow V, E$ | |

From the algorithm, we can see that the time complexity of graph construction is $O(\frac{n^2}{2})$, where $n$ is the number of data flows.

### 3.3 BS based GAT

From the algorithm of GAT, we can see that all attention coefficients are learned automatically through a neural network, where attention coefficients are similar to the similarity values between a node and its neighboring nodes. Relying solely on automatically learned attention coefficients to aggregate neighboring nodes not fully capture the information of the graph constructed in Section 3.2 of this paper. The introduction of behavior similarity is intended to fully incorporate behavior similarity into the graph neural network.

The attention coefficients, as the coefficient values between two nodes, should be constrained by behavioral similarity. The aggregation process of node hidden layer features should be influenced by edge behavior weighting. Therefore, behavior weighting is introduced into the calculation process of attention coefficients:

$$\alpha_{ij} = \frac{eb_{ij}(\exp(\text{LeakyReLU}(\vec{a}^T[W\vec{v}_i||W\vec{v}_k])))}{\sum_{k \in \mathcal{N}_i} eb_{ik}(\exp(\text{LeakyReLU}(\vec{a}^T[W\vec{v}_i||W\vec{v}_k])))} \quad (5)$$

Edge behavior weighting is introduced into the aggregation process of calculating the hidden layer features of nodes:

$$\vec{v}'_i = \sigma(\frac{1}{K}\sum_{k=1}^{K}\sum_{j \in \mathcal{N}_i} eb_{ij}\alpha_{ij}^k W^k \vec{v}_j) \quad (6)$$

Equation (5) considers the behavior weight of all neighboring nodes, while equation (6) does not take this into account. This paper adopts the method of equation (5).

Through the above methods, we write the pseudocode as follows:

---
**Algorithm 2: BS-GAT minibatch aggregation**

---
Input: Graph $\mathcal{G}(V, E)$;

        node features $\vec{v}_i$;

        layer $L$; number of heads $K$;

        weight matrices $W^k$, $\forall k \in \{1,\dots,K\}$;

        non-linearity $\sigma$;

        neighborhood for node $i$ $\mathcal{N}_i$;

        behavior weight of edge $eb_{ij}$;

Output: node features $\vec{v}_i^l$;

---

1   $\vec{v}_i^0 \leftarrow \vec{v}_i$

2   **for** $l = 1,\dots,L$ **do**

3      **for** $k = 1,\dots,K$ **do**

4      
$$\alpha_{ij}^k = \frac{eb_{ij}(\exp\left(\text{LeakyReLU}(\vec{a}^T[W^k\vec{v}_i||W^k\vec{v}_j])\right))}{\sum_{s\in\mathcal{N}_i} eb_{is}(\exp(\text{LeakyReLU}(\vec{a}^T[W^k\vec{v}_i||W^k\vec{v}_s])))}$$

5      **end**

6   
$$\vec{v}_i^l = \sigma(\frac{1}{K}\sum_{k=1}^{K}\sum_{s\in\mathcal{N}_i}\alpha_{ij}^k W^k \vec{v}_s)$$

7   **end**

---

From the algorithm, we can see that the time complexity of BS-GAT is $O(n)$, where $n$ is the number of data flows.

## 4 Experimental Evaluation

In order to evaluate the performance of the proposed method, experiments are conducted on two latest stream-based intrusion detection datasets. First, describe the data set used briefly, and then elaborate on the experimental setup, evaluation metrics and experimental methods. Finally, analyze the experimental results.

### 4.1 Data Sets

Two data sets NF-BoT-IoT-v2 and NF-ToN-IoT-v2, which are mentioned in the paper[24], are used for experiments. These two datasets are improved on the basis of the datasets NF-BoT-IoT and NF-ToN-IoT[25], and their main information is shown in Table 1.

Table 1: Information comparison between two different versions of data sets.

| Dataset | Release year | No. Classes | No. features | No. data | Benign ratio |
|---------|--------------|-------------|--------------|----------|--------------|

| Dataset | Year | | | | |
|---|---|---|---|---|---|
| NF-BoT-IoT | 2020 | 5 | 12 | 3,668,522 | 0.2 to 9.8 |
| NF-BoT-IoT-v2 | 2021 | 5 | 43 | 37,763,497 | 0.0 to 10.0 |
| NF-ToN-IoT | 2020 | 10 | 12 | 22,339,021 | 2.0 to 8.0 |
| NF-ToN-IoT-v2 | 2021 | 10 | 43 | 16,940,496 | 3.6 to 6.4 |

These data sets are re-formatted by network analysis tools on real data sets BoT-IoT[26] and ToN-IoT[27], making them more in line with the characteristics of network intrusion detection. Both datasets support multi-classification, and the amount of data in these datasets is very large and complete. The biggest change of these datasets in the version of v2 is expansion of feature fields from 12 to 43, which has a great impact on classification and identification. The multi-category information for these two datasets is also analyzed, and the results are shown in Table 3.

Table 2: Ratio of categories to total data volume in dataset

| Dataset | Classes(names and %) | | | | | | | | | |
|---|---|---|---|---|---|---|---|---|---|---|
| NF-BoT-IoT-v2 | Benign | Reconnaissance | DDos | Dos | Theft | | | | | |
| | 0.36 | 6.94 | 48.54 | 44.15 | 0.0064 | | | | | |
| NF-ToN-IoT-v2 | Benign | Reconnaissance | DDos | Dos | Backdoor | Injection | MITM | Password | Scanning | XSS |
| | 36.01 | 0.002 | 11.96 | 4.2 | 0.0099 | 4.04 | 0.0046 | 6.8 | 22.32 | 14.49 |

As these two datasets are very large, with sizes of 5.63GB and 2.49GB for the NF-BoT-IoT-v2 and NF-ToN-IoT-v2 datasets, respectively, this paper used a portion of the datasets for experimentation in order to improve efficiency. The experiment utilized all available data for the categories with relatively smaller amounts of data in the datasets, such as "Benign" and "Theft" for NF-BoT-IoT-v2, and "Reconnaissance," "Backdoor," and "MITM" for NF-ToN-IoT-v2. For the other categories, 5% of the data was randomly selected and used in the experiment. The final number of samples used in the experiment for NF-BoT-IoT-v2 dataset was 2,018,770, and for NF-ToN-IoT-v2 dataset it was 934,576.

### 4.2 Experimental Setting

**a) Evaluation Metrics**

The experiment in this paper is divided into two parts: the uniformity of node neighbors and the accuracy of classification identification. Therefore, the evaluation metrics for these two parts will be discussed in the experiment.

**(1) Average degree of the number of neighbors of a node**

This paper uses the Gini coefficient to evaluate the average degree of the number of neighbors of a node, and its calculation method and schematic diagram are shown as equation (7) and Figure 4 below:

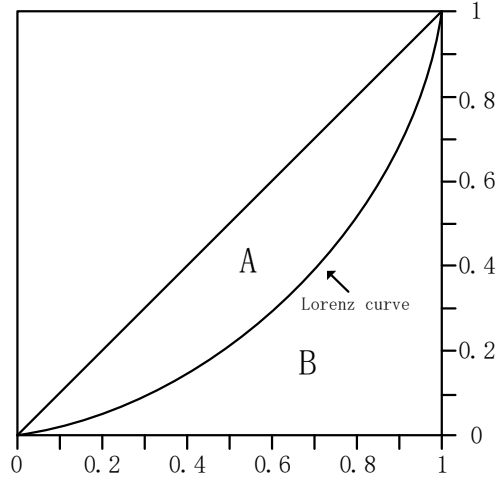

Figure 4 Gini coefficient diagram

$$G = \frac{S_A}{S_A + S_B} \tag{7}$$

Where $S_A$ is the area of region A, and $S_B$ is the area of region B. The smaller the Gini coefficient, the more evenly distributed the number of neighbors of the node.

Sort the number of neighbors for all nodes in ascending order, and calculate the percentage of the total number of neighbors for nodes in the top 10% to top 90% of nodes, fits the curve using a regression equation, then calculates the area of $S_B$ using calculus functions, and finally calculates the Gini coefficient $G$.

**(2) Classification accuracy**

In the experiment, the proposed method approach was compared with graph neural network-based methods (E-GraphSAG, E-ResGAT, E-ResGAT) and traditional machine learning methods (KNN, SVM, Random Forest). To evaluate the performance of the proposed algorithm, the evaluation metrics shown in Table 3 were used. Where TP, FP, FN, and TN represent True Positive, False Positive, True Negative, and False Negative, respectively, in the Confusion Matrix.

Table 3: Evaluation metrics adopted in this paper

| Metric | Definition |
|---|---|
| Recall | $\dfrac{TP}{TP + FN}$ |
| Precision | $\dfrac{TP}{TP + FP}$ |
| F1-Score | $2 \times \dfrac{Recall \times Precision}{Recall + Precision}$ |
| Accuracy | $\dfrac{TP + TN}{TP + FP + FN + TN}$ |

**b) Parameter setting**

In the experiment, the dataset was divided into three parts: 50% for the training set, 20% for the validation set, and 30% for the test set. The hyperparameters involved in the experiment are shown in Figure 5, and detailed hyperparameter values were provided for each algorithm. Except for the hyperparameter values specified in this paper for BS-GAT, the other algorithms used the hyperparameter values reported in their original papers.

Table 4: Experimental hyperparameter settings

| Name of parameter | BS-GAT | E-GraphSAG | E-ResGAT | GAT |
|---|---|---|---|---|
| learning rate | 0.002 | 0.001 | 0.003 | 0.002 |
| dropout | 0.2 | 0.2 | 0.2 | 0.2 |
| hop | 1 | 2 | 2 | 1 |
| hidden | 128 | 128 | 128 | 128 |
| layers | 2 | 2 | 2 | 2 |
| minibatchs | 500 | 500 | 500 | 500 |
| loss function | Cross-Entropy | Cross-Entropy | Cross-Entropy | Cross-Entropy |
| optimizer | Adam | Adam | Adam | Adam |
| epoch | 1000 | 1000 | 1000 | 1000 |
| nonlinear function | eLU | ReLU | ReLU | eLU |
| nb_heads | 3 | — | 6 | 3 |
| $\lambda$ | 0.85 | — | — | — |
| $\mu$ | 0.7 | — | — | — |

Using the hyperparameters shown in Table 4, the final feature values of each node in the last layer of the graph neural network were obtained, and then these values were fed into a fully connected layer. Softmax was used for multi-class classification after the fully connected layer.

### 4.3 Results

a) Gini coefficient

This paper has statistically analyzed the Gini coefficients of the graphs generated by three methods, and the results are shown in Table 5.

Table 5: Statistical results of the Gini coefficient.

| | Gini coefficient | |
|---|---|---|
| | NF-BoT-IoT-v2 | NF-ToN-IoT-v2 |
| BS-GAT | **0.3529** | **0.4203** |
| E-GraphSAGE | **0.8845** | **0.9216** |
| E-ResGAT | **0.8903** | **0.9176** |

The results in Table 5 show that the proposed graph construction method has a smaller Gini coefficient than the other two methods. The graphs generated by the proposed method are more average, and using graph neural networks for identification will lead to better results.

b) Classification results

Because there are two labels, Label and Attack, in the datasets, where Label indicates whether the network flow data is an attack or not, and Attack label defines the specific type of the network data flow. Therefore, binary classification and multi-class classification can be performed using both Label and Attack labels for prediction.

**(1) Binary classification results**

First, we conducted experiments on binary classification. Since only a portion of the dataset was used for experimentation, multiple experiments were performed for each algorithm and dataset, and the average results were recorded, as shown in Table 6.

Table 6: Results of each algorithm on binary classification

| DataSet | Algorithm | Recall | Precision | F1-Score | Accuracy |
| --- | --- | --- | --- | --- | --- |
| NF-BoT-IoT-v2 | KNN | 0.9425 | 0.9217 | 0.9319 | 0.9259 |
| | SVM | 0.9602 | 0.9543 | 0.9572 | 0.9198 |
| | Random Forest | 0.9489 | 0.9125 | 0.9303 | 0.9315 |
| | GAT | 0.9747 | 0.9931 | 0.9838 | 0.984 |
| | E-GraphSAG | 0.9831 | 0.9461 | 0.9642 | 0.9635 |
| | E-ResGAT | 0.9621 | 0.92 | 0.9406 | 0.9392 |
| | BS-GAT | **0.9835** | **0.9964** | **0.9899** | **0.99** |
| NF-ToN-IoT-v2 | KNN | 0.9272 | 0.9147 | 0.9209 | 0.8904 |
| | SVM | 0.9526 | 0.9248 | 0.9384 | 0.9206 |
| | Random Forest | 0.9458 | 0.9648 | 0.9552 | 0.9368 |
| | GAT | 0.9845 | 0.9556 | 0.9420 | 0.9694 |
| | E-GraphSAG | 0.9705 | 0.9339 | 0.9518 | 0.9509 |
| | E-ResGAT | 0.977 | **0.9696** | 0.9735 | 0.9734 |
| | BS-GAT | **0.9922** | 0.9663 | **0.9790** | **0.9788** |

From Table 6, we can see that all seven algorithms achieved good results in the Recall metric for both datasets in the binary classification problem, with all achieving above 94%. This is consistent with the findings of the E-GraphSAG and E-ResGAT papers, which also reported good results in binary classification. However, the algorithm proposed in this paper still outperformed the other six algorithms on both datasets. In addition, the algorithm proposed in this paper also has a good performance on the Precision. The result is only 0.33 % lower than E-ResGAT on the data set NF-ToN-IoT-v2, but 1.07 % and 3.24 % higher than GAT and E-GraphSAG, respectively. In addition, the algorithm proposed in this paper is better than the other six algorithms on the F1-Score, showing that it has good stability.

**(2) Multiclass classification results**

The experimental method of the multi-classification problem is the same as the binary classification problem, and the average value is taken after multiple experiments. Finally, the multi-classification results of each algorithm for different data sets are shown in table 7 and the overall performance of each algorithm for multi-classification is shown in table 8.

Table 7: Recall results per class on the tow datasets.

| Dataset | Algorithm | Per class Recall | | | | | | | | | |
|---|---|---|---|---|---|---|---|---|---|---|---|
| | | Benign | Reconnaissance | DDos | Dos | Theft | | | | | |
| NF-BoT-IoT-v2 | KNN | 0.8419 | 0.8754 | 0.8025 | 0.7136 | 0.8526 | | | | | |
| | SVM | 0.8615 | 0.7256 | 0.7259 | 0.8649 | 0.7215 | | | | | |
| | Random Forest | 0.9105 | 0.9025 | 0.8425 | 0.8926 | 0.9145 | | | | | |
| | GAT | 0.9712 | 0.8426 | 0.8122 | 0.8929 | 0.9517 | | | | | |
| | E-GraphSAG | 0.9916 | 0.925 | 0.8382 | 0.9524 | **0.9965** | | | | | |
| | E-ResGAT | 0.9962 | 0.9527 | 0.9693 | 0.9119 | 0.9326 | | | | | |
| | BS-GAT | **0.9984** | **0.9811** | **0.9725** | **0.9876** | 0.9803 | | | | | |
| | | Benign | Reconnaissance | DDos | Dos | Backdoor | Injection | MITM | Password | Scanning | XSS |
| NF-ToN-IoT-v2 | KNN | 0.8136 | 0.7845 | 0.7652 | 0.8695 | 0.6584 | 0.8456 | 0.7458 | 0.8016 | 0.8475 | 0.7852 |
| | SVM | 0.8514 | 0.8268 | 0.7485 | 0.6254 | 0.7854 | 0.8912 | 0.7962 | 0.9025 | 0.7958 | 0.8456 |
| | Random Forest | 0.8054 | 0.9258 | 0.7469 | 0.8549 | 0.7608 | 0.8647 | 0.6058 | 0.8926 | 0.8746 | 0.8106 |
| | GAT | 0.8512 | 0.9024 | 0.8563 | 0.8015 | 0.9104 | 0.8205 | 0.552 | 0.7425 | 0.8984 | 0.9156 |
| | E-GraphSAG | 0.8943 | 0.9634 | 0.8947 | 0.844 | 0.9811 | 0.9328 | 0.7651 | 0.9023 | 0.8042 | 0.925 |
| | E-ResGAT | 0.8896 | 0.9457 | **0.9228** | **0.906** | 0.9226 | 0.8825 | 0.9044 | 0.9419 | 0.7831 | 0.9639 |
| | BS-GAT | **0.9214** | **0.9951** | 0.9162 | 0.8574 | **0.9892** | **0.9959** | **0.9821** | **0.9614** | **0.9155** | **0.9956** |

From Table 7, it can be seen that the proposed algorithm in this paper has superiority in multi-classification problems, with high identification accuracy for each classification in different datasets. On the NF-BoT-IoT-v2 dataset, the proposed algorithm has a 3.47% - 20.78% improvement in Weighted Recall and a 4.16% - 15.55% improvement in Weighted F1-Score compared to other algorithms. On the NF-ToN-IoT-v2 dataset, the proposed algorithm has a 4.82% - 12.36% improvement in Weighted Recall and a 4.98% - 10.42% improvement in Weighted F1-Score compared to other algorithms. From the experimental results, it can be seen that E-GraphSAG has also improved in the multi-classification problem compared to its original paper, which used the v1 version of the dataset. This indicates that the dataset used in this paper has been greatly improved compared to the first version, demonstrating the correctness of the dataset used.

Table 8: Performance results for each algorithm over multiple classifications

| DataSet | Algorithm | Weighted Recall | Weighted F1-Score |
|---|---|---|---|
| NF-BoT-IoT-v2 | KNN | 0.7732 | 0.8192 |
| | SVM | 0.7923 | 0.831 |
| | Random Forest | 0.8717 | 0.8777 |
| | GAT | 0.8583 | 0.8701 |
| | E-GraphSAG | 0.9014 | 0.9007 |
| | E-ResGAT | 0.9463 | 0.9331 |
| | BS-GAT | **0.981** | **0.9747** |
| NF-ToN-IoT-v2 | KNN | 0.8105 | 0.848 |
| | SVM | 0.8224 | 0.8552 |
| | Random Forest | 0.8211 | 0.8544 |
| | GAT | 0.8486 | 0.8705 |
| | E-GraphSAG | 0.880 | 0.9024 |
| | E-ResGAT | 0.8859 | 0.8966 |
| | BS-GAT | **0.9341** | **0.9522** |

The superiority of the method proposed in this paper in the problem of network intrusion detection, especially in multi-classification problems, has been demonstrated through the above binary and multi-classification experiments.

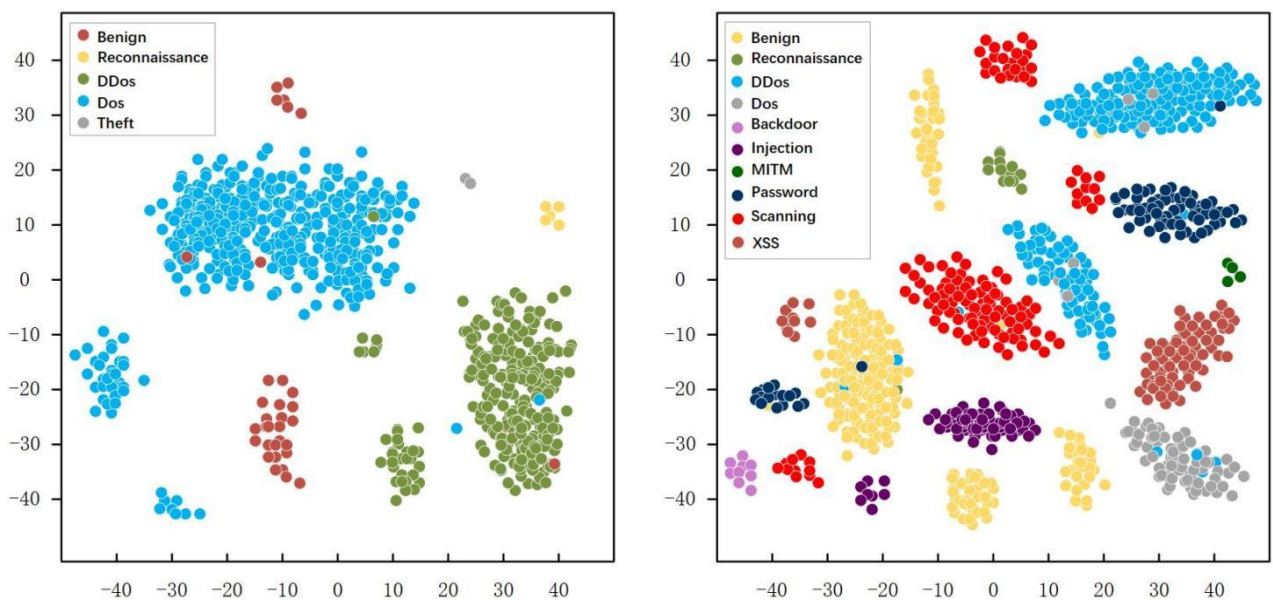

Figure 4: The features of the hidden layer finally obtained from the BS-GAT model are displayed using the t-SEN visualization method. **Left:** Dataset of NF-BoT-IoT-v2. **Right:** Dataset of NF-ToN-IoT-v2.

Finally, we used the t-SEN method to visually display the node characteristics of the last layer of BS-GAT. T-SNE is a method of transforming multidimensional features into two-dimensional features and visualizing them. Through t-SNE, we can see the clustering results of the BS-GAT model. This paper shows the visualization effects of the two data sets shown in Figure 4.

## 5 Conclusions

In this paper, we proposed a graph attention network (BS-GAT) based on behavioral similarity and applied it to network intrusion detection. A novel graph construction method is proposed by using behavioral similarity, where the data flows are treated as nodes in the graph, and the behavior rules of nodes are used as edges in the graph. A behavioral weight matrix is constructed between nodes, which is applied to GAT for utilizing the relationship between data flows and the structure information of the graph. Finally, experiments are conducted using the latest flow-based network intrusion detection dataset. Comparisons are made between the proposed method and existing methods in order to demonstrate its superiority and feasibility.